%% file: tauv_rhov_v2.tex
\documentclass[aps,prd,twocolumn,showpacs,amsmath,amssymb]{revtex4-1}
\usepackage{amsmath}
\usepackage{graphicx}
\usepackage{subfigure}
\usepackage{epstopdf}
\usepackage{color}
\usepackage{multirow}
\usepackage{setspace}
\usepackage{overpic}
\usepackage{amssymb}
\usepackage{makecell}

\usepackage[colorlinks,urlcolor=blue, citecolor=blue,linkcolor=blue] {hyperref}
\usepackage{lineno}
\usepackage{bm}
\usepackage{rotating}
\usepackage[utf8]{inputenc}

\let\oldequation\equation
\let\oldendequation\endequation
\renewenvironment{equation}
  {\linenomathNonumbers\oldequation}
  {\oldendequation\endlinenomath}

\begin{document}

\title{\bf \boldmath
Measurement of the branching fraction of leptonic decay $D_s^+\to\tau^+\nu_\tau$ via $\tau^+\to\pi^+\pi^0\bar \nu_\tau$
}

\input{BESIII_author}

\linenumbers

\begin{abstract}
By analyzing $6.32~\mathrm{fb}^{-1}$ of $e^+e^-$ annihilation data collected at the center-of-mass energies between 4.178 and 4.226\,GeV with the BESIII detector,
we determine the branching fraction of the leptonic decay $D_s^+\to\tau^+\nu_\tau$ with $\tau^+\to\pi^+\pi^0\bar \nu_\tau$,
to be
$\mathcal{B}_{D_s^+\to\tau^+\nu_\tau}=(5.29\pm0.25_{\rm stat}\pm0.20_{\rm syst})\%$.
We estimate the product of the Cabibbo-Kobayashi-Maskawa matrix element $|V_{cs}|$ and the $D_s^+$ decay constant $f_{D^+_s}$ to be $f_{D_s^+}|V_{cs}|=(244.8\pm5.8_{\rm stat}\pm4.8_{\rm syst})~\mathrm{MeV}$
using the known values of the $\tau^+$ and $D_s^+$ masses as well as the $D_s^+$ lifetime, together with our branching fraction measurement. Combining with the value of $|V_{cs}|$ obtained from a global fit in the standard model and $f_{D_s^+}$ from lattice quantum chromodynamics, we obtain $f_{D_s^+}=(251.6\pm5.9_{\rm stat}\pm4.9_{\rm syst})$\,MeV and $|V_{cs}| = 0.980\pm0.023_{\rm stat}\pm0.019_{\rm syst}$. Using the branching fraction of $\mathcal B_{D_s^+\to\mu^+\nu_\mu}=(5.35\pm0.21)\times10^{-3}$, we obtain the ratio of the branching fractions $\mathcal B_{D_s^+\to\tau^+\nu_\tau}/\mathcal B_{D_s^+\to\mu^+\nu_\mu}=9.89\pm0.71$, which is consistent with the standard model prediction of lepton flavor universality.
\end{abstract}


\maketitle

\oddsidemargin  -0.2cm
\evensidemargin -0.2cm

\section{Introduction}
In the standard model, the partial width for the leptonic decay $D^+_s\to \ell^+\nu_\ell$~($\ell=e$, $\mu$ or $\tau$)
is written as~\cite{decayrate}
\begin{equation}
\Gamma_{D^+_{s}\to\ell^+\nu_\ell}=\frac{G_F^2}{8\pi}|V_{cs}|^2
f^2_{D^+_{s}}
m_\ell^2 m_{D^+_{s}} \left (1-\frac{m_\ell^2}{m_{D^+_{s}}^2} \right )^2,
\end{equation}
where
$f_{D^+_{s}}$ is the $D^+_{s}$ decay constant,
$|V_{cs}|$ is the
Cabibbo-Kobayashi-Maskawa~(CKM) matrix element describing the relative strength of $c$ quark to $s$ quark transition,
$G_F$ is the Fermi coupling constant,
$m_\ell$ is the lepton mass, and
$m_{D^+_{s}}$ is the $D^+_{s}$ mass.
Charge conjugations are always included throughout this paper.
The $D^+_s\to \ell^+\nu_\ell$ decays offer an ideal opportunity to determine
 $f_{D^+_{s}}$ or $|V_{cs}|$ in case the other has been given.
Previously, the
CLEO~\cite{cleo2009,cleo2009a,cleo2009b}, BaBar~\cite{babar2010}, Belle~\cite{belle2013}, and
BESIII~\cite{bes2016,bes2019,hajime2021} collaborations have reported the measurements of the $D^+_s\to \ell^+\nu_\ell$ decays, giving an averaged precision for $f_{D^+_s}$ of 1.5\%.
In contrast, $f_{D^+_s}$ has been well calculated by Lattice Quantum Chromodynamics~(LQCD) with an uncertainty of
0.2\%~\cite{FLab2018}.
Improved measurements of $f_{D^+_{s}}$ in experiment are important to test various theoretical calculations~\cite{FLab2018,LQCD,etm2015,ukqcd2017,ukqcd2015,FLAG2019,chen2014,becirevic2013,wang2015}. Meanwhile, precise measurements of $|V_{cs}|$ are also important to test the CKM matrix unitarity~\cite{arxiv210300908}.

On the other hand, the ratio of the branching fractions of $D^+_s\to \tau^+\nu_\tau$
and $D^+_s\to \mu^+\nu_\mu$,
\begin{equation}
\mathcal R_{\tau/\mu}=\frac{\mathcal B_{D^+_s\to \tau^+\nu_\tau}}{\mathcal B_{D^+_s\to \mu^+\nu_\mu}}=\frac{m_{\tau^+}^2(1-\frac{m_{\tau^+}^2}{m_{D^+_s}^2})^2}{m_{\mu^+}^2(1-\frac{m_{\mu^+}^2}{m_{D^+_s}^2})^2},
\end{equation}
 in the standard model with the implication of lepton flavor universality predicts to be 9.75$\pm$0.01 using the world averages of $m_\tau$, $m_\mu$, and $m_{D_s}$~\cite{PDG2020}.
In the BaBar, LHCb, and Belle experiments, however, hints of lepton flavor universality violation in semileptonic $B$ decays have been reported in recent years~\cite{babar_1,babar_2,lhcb_1,belle2015,belle2016,lhcb_1a,belle2019}.
Examination of lepton flavor universality in the $D^+_s\to\ell^+\nu_\ell$ decays is therefore important to test lepton flavor universality.

This paper reports a measurement of the branching fraction for $D_s^+\to\tau^+\nu_\tau$
via $\tau^+\to\pi^+\pi^0\bar \nu_\tau$. This analysis is performed by using the data samples collected at the center-of-mass energies
$\sqrt s=4.178$, 4.189, 4.199, 4.209, 4.219, and 4.226\,GeV with the BESIII detector.
The total integrated luminosity of these data samples is $6.32~\mathrm{fb}^{-1}$.

\section{BESIII detector and Monte Carlo simulations}

The BESIII detector~\cite{Ablikim:2009aa} records symmetric $e^+e^-$ collisions
provided by the BEPCII storage ring~\cite{Yu:IPAC2016-TUYA01}, which operates with a peak luminosity of $1\times10^{33}$~cm$^{-2}$s$^{-1}$
in the center-of-mass energy range from 2.0 to 4.95~GeV.
BESIII has collected large data samples in this energy region~\cite{Ablikim:2019hff}. The cylindrical core of the BESIII detector covers 93\% of the full solid angle and consists of a helium-based
 multilayer drift chamber~(MDC), a plastic scintillator time-of-flight
system~(TOF), and a CsI(Tl) electromagnetic calorimeter~(EMC),
which are all enclosed in a superconducting solenoidal magnet
providing a 1.0~T magnetic field. The solenoid is supported by an
octagonal flux-return yoke with resistive plate counter muon
identification modules interleaved with steel.
The charged-particle momentum resolution at $1~{\rm GeV}/c$ is
$0.5\%$, and the $dE/dx$ resolution is $6\%$ for electrons
from Bhabha scattering. The EMC measures photon energies with a
resolution of $2.5\%$ ($5\%$) at $1$~GeV in the barrel (end cap)
region. The time resolution in the TOF barrel region is 68~ps. The end cap TOF
system was upgraded in 2015 using multi-gap resistive plate chamber
technology, providing a time resolution of
60~ps~\cite{etof}.

Simulated data samples produced with a {\sc
geant4}-based~\cite{geant4} Monte Carlo (MC) package, which
includes the geometric description of the BESIII detector and the
detector response, are used to determine detection efficiencies
and to estimate backgrounds. The simulation models the beam
energy spread and initial state radiation (ISR) in the $e^+e^-$
annihilations with the generator {\sc
kkmc}~\cite{ref:kkmc}.
In the simulation, the production of open-charm
processes directly produced via $e^+e^-$ annihilations are modeled with the generator {\sc conexc}~\cite{ref:conexc},
and their subsequent decays are modeled by {\sc evtgen}~\cite{ref:evtgen} with
known branching fractions from the Particle Data Group~\cite{pdg}.
The ISR production of vector charmonium(-like) states
and the continuum processes are incorporated in {\sc
kkmc}~\cite{ref:kkmc}.
The remaining unknown charmonium decays
are modelled with {\sc lundcharm}~\cite{ref:lundcharm}. Final state radiation
from charged final-state particles is incorporated using the {\sc
photos} package~\cite{photos}.

\section{Analysis method}
Similar double-tag (DT) method used in Refs.~\cite{hajime2021,DTmethod} is employed in this article,
 At $\sqrt s$ between 4.178 and 4.226~GeV, $D_s^+$ mesons are produced mainly from the processes $e^+e^-\to D_s^{*\pm}[\to\gamma(\pi^0)D_s^\pm]D_s^\mp$. We first fully reconstruct one $D_s^-$ meson in one of several hadronic decay modes, called as a single-tag candidate. We then examine the signal decay of the $D_s^+$ meson and the $\gamma(\pi^0)$ from $D_s^{*+}$, named as a double-tag candidate.
At the $j$-th energy point, $j$=0, 1, 2, 3, 4, and 5 for the energy points 4.178, 4.189, 4.199, 4.209, 4.219, and 4.226 GeV, respectively, the branching fraction for $D^+_s\to \tau^+\nu_\tau$ is determined by
\begin{equation}
\mathcal B_{D_s^+\to\tau^+\nu_\tau}=\frac{N_{\rm DT}^j}{N_{\rm ST}^{j} \cdot \epsilon^j_{\gamma(\pi^0)\tau^+\nu_\tau}\cdot\mathcal B_{\rm sub}}.
\label{eq1}
\end{equation}
Here, $N_{\rm DT}^j$ is the double-tag yield in data; $N_{\rm ST}^j=\Sigma_i N_{\rm ST}^{ij}$ is the total single-tag yield in data summing over tag mode $i$;
$\epsilon^j_{\gamma(\pi^0)\tau^+\nu_\tau}$ is the efficiency of detecting $D^+_s\to \tau^+\nu_\tau$ in the presence of the single-tag $D^-_s$ candidate, averaged by the single-tag yields in data.
It is calculated by $\Sigma_i(N_{\rm ST}^{ij}/N_{\rm ST}^{j}) \cdot (\epsilon^{ij}_{\rm DT}/\epsilon^{ij}_{\rm ST})$, where $\epsilon^{ij}_{\rm DT}$ and $\epsilon^{ij}_{\rm ST}$ are
the detection efficiencies of the double-tag and single-tag candidates, respectively. The efficiencies do not include the branching fractions for the sub-resonant decays. $\mathcal B_{\rm sub}$ is the product of the branching fractions for the $\tau^+\to\pi^+\pi^0\bar{\nu}_\tau$ and $\pi^0\to\gamma\gamma$ decays.

\section{Single-tag $D^-_s$ candidates}

The single-tag $D^-_s$ candidates are reconstructed from fourteen hadronic decay modes of
$D^-_s\to K^+K^-\pi^-$, $K^+K^-\pi^-\pi^0$, $K^0_SK^-$,
$K^0_SK^-\pi^0$,
$K^0_SK^0_S\pi^-$,
$K^0_SK^+\pi^-\pi^-$,
$K^0_SK^-\pi^+\pi^-$,
$\pi^+\pi^-\pi^-$,
$\eta_{\gamma\gamma}\pi^-$, $\eta_{\pi^0\pi^+\pi^-}\pi^-$,
$\eta^\prime_{\eta_{\gamma\gamma}\pi^+\pi^-}\pi^-$, $\eta^\prime_{\gamma\rho^0}\pi^-$,
$\eta_{\gamma\gamma}\rho^-$, and
$\eta_{\pi^+\pi^-\pi^0}\rho^-$,
where the subscripts of $\eta$ and $\eta^{\prime}$ represent the decay modes used to reconstruct $\eta$ and $\eta^{\prime}$, respectively. Throughout this paper, $\rho$ denotes $\rho(770)$.

The selection criteria of $K^\pm$, $\pi^\pm$, $K^0_S$, $\gamma$, $\pi^0$, and $\eta$ are the same as those used in our previous works~\cite{bes2019,bes3_etaev,bes3_gev}.
All charged tracks must satisfy
$|V_{xy}|<1$ cm, $|V_{z}|<10$ cm, and $|\!\cos\theta|<0.93$,
where $|V_{xy}|$ and $|V_{z}|$ are a distance of the closest approach in the transverse plane and
along the MDC axis, respectively, and $\theta$ is the polar angle with respect to the MDC axis. This requirement is not applied for those from $K_S^0$ decays.
Particle identification (PID) of the charged particles is performed with the combined $dE/dx$ and TOF information.
The confidence levels for pion and kaon hypotheses ($CL_\pi$ and $CL_K$) are obtained.
Kaon and pion candidates are required to satisfy $CL_{K}>CL_{\pi}$ and $CL_{\pi}>CL_{K}$, respectively.

The $K_S^0$ mesons are reconstructed via the $K^0_S\to \pi^+\pi^-$ decays.
The distances of the closest approach of the two charged pions to the interaction point are required to be less than 20 cm along the MDC axis.
They are assumed to be $\pi^+\pi^-$ without PID requirements.
The invariant mass of the $\pi^+\pi^-$ combination is required to be within $\pm12$\,MeV$/c^2$ around the $K^0_S$ nominal mass~\cite{PDG2020}.
The decay length of the reconstructed $K_S^0$ is required to be greater than twice of the vertex resolution away from the interaction point.

The $\pi^0$ and $\eta$ mesons are reconstructed from photon pairs.
Photon candidates are selected from the shower clusters in the EMC that are not associated with a charged track.
Each electromagnetic shower is required to start within 700\,ns of the event start time.
The shower energy is required to be greater than 25\,(50)\,MeV
in the barrel\,(end cap) region of the EMC~\cite{Ablikim:2009aa}.
The opening angle between the candidate shower and
the nearest charged track is required to be greater than $10^{\circ}$.
To form $\pi^0$ and $\eta$ candidates, the invariant masses of the selected photon pairs are required to be
within the $M_{\gamma\gamma}$ interval $(0.115,\,0.150)$ and $(0.50,\,0.57)$\,GeV$/c^{2}$, respectively.
To improve momentum resolution and suppress background,
a kinematic fit is imposed on each chosen photon pair to constrain its invariant mass to the $\pi^{0}$ or $\eta$ nominal mass~\cite{PDG2020}.

For the tag modes $D^-_s\to \eta\pi^-$ and $\eta\rho^-$,
the $\pi^0\pi^+\pi^-$ combinations used to form $\eta$ candidates are required to be
within the $M_{\pi^0\pi^+\pi^-}$ interval $(0.53,\,0.57)~\mathrm{GeV}/c^2$.
To form $\eta^\prime$ candidates, we use two decay modes $\eta\pi^+\pi^-$ and $\gamma\rho^0$,
whose invariant masses are required to be within
the interval $(0.946,\,0.970)$ GeV/$c^2$ and $(0.940,\,0.976)~\mathrm{GeV}/c^2$, respectively.
In addition, the minimum energy
of the $\gamma$ from $\eta'\to\gamma\rho^0$ decays must be greater than 0.1\,GeV.
The $\rho^0$ and $\rho^+$ candidates are reconstructed from the $\pi^+\pi^-$ and $\pi^+\pi^0$
combinations with invariant masses within the interval $(0.57,\,0.97)~\mathrm{GeV}/c^2$.

To reject the soft pions from $D^{*+}$ decays,
the momentum of any pion, which does not originate from a $K_S^0$, $\eta$, or $\eta^\prime$ decay, is required to be greater than 0.1\,GeV/$c$.
For the tag mode $D^-_s\to \pi^+\pi^-\pi^-$, the peaking background from $D^-_s\to K^0_S\pi^-$ final state is rejected by requiring any $\pi^+\pi^-$ combination to be outside of the mass window $\pm 0.03$ GeV/$c^2$ around the $K^0_S$ nominal mass~\cite{PDG2020}.

To suppress non $D_s^{\pm}D^{*\mp}_s$ events,
the beam-constrained mass of the single-tag $D_s^-$
candidate
\begin{equation}
M_{\rm BC}\equiv\sqrt{E^2_{\rm beam}-|\vec{p}_{\rm tag}|^2}
\end{equation}
is required to be within $(2.010,\,2.073+j\times0.003)\,\mathrm{GeV}/c^2$,
where $E_{\rm beam}$ is the beam energy and
$\vec{p}_{\rm tag}$ is the momentum of the single-tag $D_s^-$ candidate in the rest frame of the initial $e^+e^-$ beams.
This requirement retains most of the $D_s^-$ mesons from $e^+ e^- \to D_s^{\pm}D_s^{*\mp}$.

\begin{figure*}[htbp]
\centering
\includegraphics[width=0.98\textwidth]
{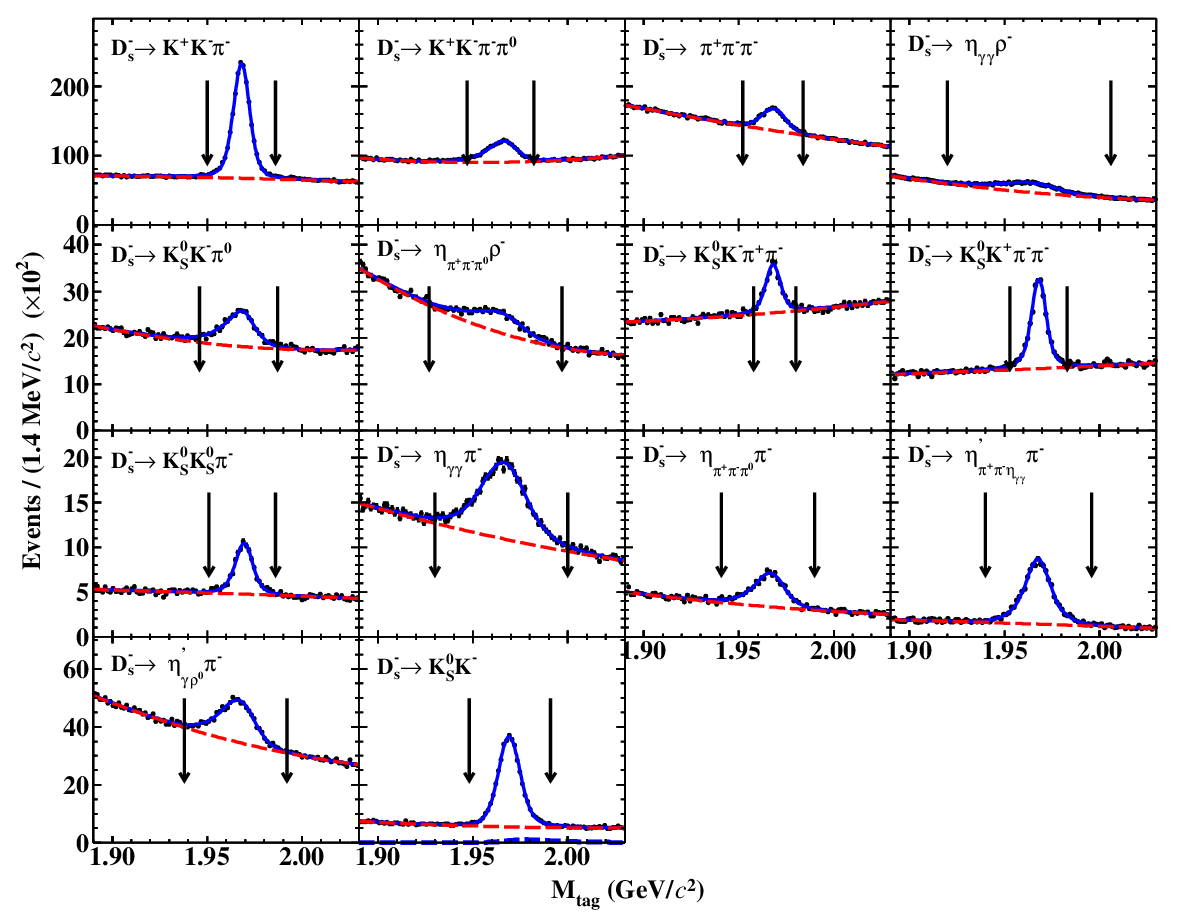}
\caption{\footnotesize
Fits to the $M_{\rm tag}$ distributions of the accepted single-tag candidates from the data sample at $\sqrt s=4.178$ GeV.
Points with error bars are data.
Blue solid curves are the fit results.
Red dashed curves are the fitted backgrounds.
Blue dotted curve in the $K_S^0K^-$ mode is the $D^-\to K_S^0\pi^-$ component.
In each sub-figure, the pair of arrows denote the signal regions.
}
\label{fig:stfit}
\end{figure*}

In each event, we only keep
one candidate with the $D_s^-$ recoil mass
\begin{equation}
M_{\rm rec} \equiv \sqrt{ \left (\sqrt s - \sqrt{|\vec p_{\rm tag}|^2+m^2_{D^-_s}} \right )^2
-|\vec p_{\rm tag}|^2}
\end{equation}
closest to the $D_s^{*+}$ nominal mass~\cite{PDG2020} per tag mode per charge.
Figure~\ref{fig:stfit} shows the invariant mass ($M_{\rm tag}$) spectra of the accepted single-tag candidates for various tag modes.
For each tag mode, the single-tag yield is obtained by a fit to
the corresponding $M_{\rm tag}$ spectrum.
The signal is described by the simulated shape convolved with a Gaussian function
representing the difference in resolution between data and simulation.
For the tag mode $D^-_s\to K_S^0K^-$,
the peaking background from $D^-\to K^0_S\pi^-$ is described by the simulated shape convolved with the same Gaussian function used in the signal shape and its size is left as a free parameter.
The non-peaking background is modeled by a first- or second-order Chebychev polynomial
function, which has been validated by using the inclusive simulation sample.
The resultant fit results for the data sample taken at $\sqrt s=4.178$ GeV are shown in Fig.~\ref{fig:stfit}. The candidates in the signal regions, denoted as the black arrows in each sub-figure, are kept for further analysis. The backgrounds from $e^+e^-\to(\gamma_{\rm ISR})D_s^+D_s^-$, which contribute about (0.7-1.1)\% in the fitted single-tag yields for various tag modes based on  simulation, are subtracted in this analysis.
As an example, the resulting single-tag yields ($N_{\rm ST}^{i1}$) for various tag modes in data at $\sqrt{s}=4.178$~GeV and the corresponding single-tag efficiencies ($\epsilon_{\rm ST}^{i1}$) are summarized in the second and third columns of Table~\ref{tab:bf}, respectively.  The individual numbers of $N^{ij}_{\rm ST}$ and $\epsilon_{\rm ST}^{ij}$ at the other energy points are obtained similarly. The total single-tag yields $N^j_{\rm ST}$ at various energy points are summarized in the second column of Table~\ref{tab:eff2}.

\begin{table}[htbp]
\centering\linespread{1.15}
        \caption{
        The obtained values of $N_{\rm ST}^{i1}$, $\epsilon_{\rm ST}^{i1}$, and $\epsilon_{\rm DT}^{i1}$ in the $i$-th tag mode at $\sqrt s=4.178$~GeV, where the efficiencies do not include the branching fractions for the sub-resonant decays and the uncertainties are statistical only. The differences among the ratios of $\epsilon_{\rm DT}^{i1}$ over $\epsilon_{\rm ST}^{i1}$ for various modes are mainly due to the requirement of $E^{\rm sum}_{\rm extra\,\gamma}$.}
\small
        \label{tab:bf}
        \begin{tabular}{l|r@{}lr@{}lr@{}l}\hline
Tag mode  &\multicolumn{2}{c}{$N_{\rm ST}^{i1}$~($\times 10^3$)}&\multicolumn{2}{c}{$\epsilon_{\rm ST}^{i1}$~(\%)}&\multicolumn{2}{c}{$\epsilon_{\rm DT}^{i1}~(\%)$} \\ \hline
$K^{+}K^{-}\pi^{-}$&137.3&$\pm$0.6&40.90&$\pm$0.04&6.80&$\pm$0.04\\
$K^{+}K^{-}\pi^{-}\pi^{0}$&42.7&$\pm$0.9&11.81&$\pm$0.04&1.75&$\pm$0.02\\
$\pi^{+}\pi^{-}\pi^{-}$&36.4&$\pm$0.9&52.12&$\pm$0.21&11.87&$\pm$0.11\\
$K_{S}^{0}K^{-}$&32.4&$\pm$0.3&49.73&$\pm$0.09&10.69&$\pm$0.11\\
$K_{S}^{0}K^{-}\pi^{0}$&11.4&$\pm$0.3&17.07&$\pm$0.13&3.60&$\pm$0.07\\
$K_{S}^{0}K_{S}^{0}\pi^{-}$&5.1&$\pm$0.1&22.77&$\pm$0.14&4.55&$\pm$0.12\\
$K_{S}^{0}K^{+}\pi^{-}\pi^{-}$&14.8&$\pm$0.2&21.05&$\pm$0.07&3.54&$\pm$0.06\\
$K_{S}^{0}K^{-}\pi^{+}\pi^{-}$&7.6&$\pm$0.3&18.47&$\pm$0.14&3.27&$\pm$0.08\\
$\eta_{\gamma\gamma}\pi^{-}$&19.4&$\pm$0.9&48.96&$\pm$0.21&10.57&$\pm$0.14\\
$\eta_{\pi^{+}\pi^{-}\pi^{0}}\pi^{-}$&5.7&$\pm$0.2&24.29&$\pm$0.16&5.61&$\pm$0.13\\
$\eta^{'}_{\pi^{+}\pi^{-}\eta_{\gamma\gamma}}\pi^{-}$&9.8&$\pm$0.1&25.43&$\pm$0.09&5.35&$\pm$0.10\\
$\eta^{'}_{\gamma\rho^{0}}\pi^{-}$&24.6&$\pm$0.7&32.51&$\pm$0.17&7.12&$\pm$0.09\\
$\eta_{\gamma\gamma}\rho^{-}$&40.8&$\pm$1.8&20.00&$\pm$0.11&4.33&$\pm$0.04\\
$\eta_{\pi^{+}\pi^{-}\pi^{0}}\rho^{-}$&11.0&$\pm$0.9&9.48&$\pm$0.11&2.07&$\pm$0.04\\
\hline
        \end{tabular}
\end{table}

\begin{table}[htbp]\centering\linespread{1.15}
	\caption{
	The total single-tag yields ($N_{\rm ST}^j$) and the averaged signal efficiencies ($\epsilon^j_{\gamma(\pi^0)\tau^+\nu_\tau}$) at various energy points, where the efficiencies do not include the branching fractions for the sub-resonant decays and the uncertainties are statistical only.}
\small
	\label{tab:eff2}
	\begin{tabular}{l|r@{}lr@{}l}\hline
$\sqrt{s}$ (GeV) &\multicolumn{2}{c}{$N_{\rm ST}^j$~($\times 10^3$)}&\multicolumn{2}{c}{$\epsilon^j_{\gamma(\pi^0)\tau^+\nu_\tau}~(\%)$}  \\ \hline
4.178&398.8&$\pm$2.8&19.01&$\pm$0.06\\
4.189&61.4&$\pm$0.8&18.55&$\pm$0.14\\
4.199&61.4&$\pm$1.0&18.43&$\pm$0.15\\
4.209&57.5&$\pm$1.0&17.77&$\pm$0.14\\
4.219&47.9&$\pm$1.1&17.24&$\pm$0.15\\
4.226&80.8&$\pm$1.6&17.19&$\pm$0.14\\
		\hline
	\end{tabular}
\end{table}

\section{Selection of $D_s^+\to \tau^+\nu_\tau$}

\begin{figure*}[htp]
  \centering
  \includegraphics[width=1\textwidth]{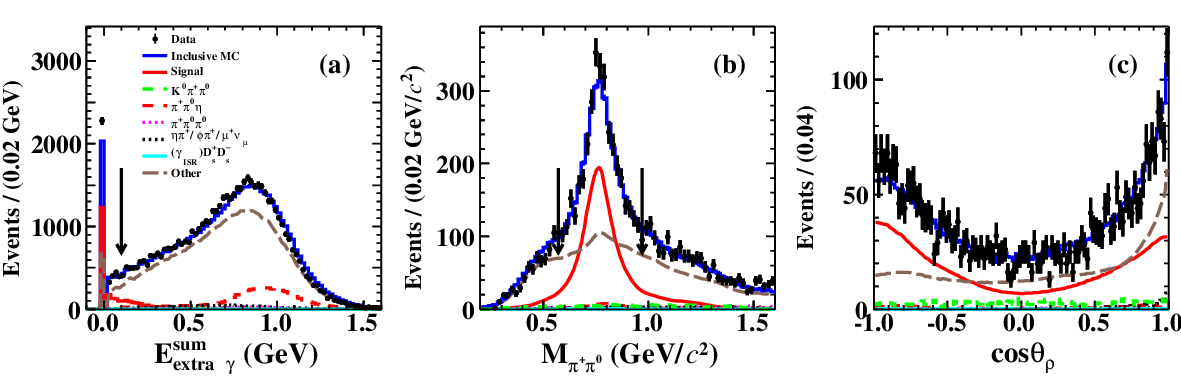}
  \caption{\footnotesize
Distributions of (a) $E_{\rm extra~\gamma}^{\rm sum}$, (b) $M_{\pi^+\pi^0}$, and (c) $\cos\theta_\rho$ of the selected $D^+\to \tau^+\nu_\tau$ candidates summed over all tag modes from all data samples.  Points with error bars are data. Blue solid lines are obtained from inclusive MC sample. Red solid lines show the signals.
Green dashed, red dashed, pink dotted, black dotted, cyan solid, and brown dashed lines are the backgrounds from $D_s^+\to K^0\pi^+\pi^0$, $D_s^+\to\pi^+\pi^0\eta$, $D_s^+\to\pi^+\pi^0\pi^0$, $D^+_s\to (\eta\pi^+,\phi\pi^+,\mu^+\nu_\mu)$, $e^+e^-\to (\gamma_{\rm ISR})D_s^+D_s^-$,
and the other backgrounds after excluding the components aforementioned, respectively. In (a) and (b), the arrows show the corresponding requirements and the events are imposed with all requirements except for the one to be shown.
}
\label{fig:fig3}
 \end{figure*}

From the recoil of the single-tag $D_s^-$ mesons,
the candidates for $D_s^+\to\tau^+\nu_\tau$ are selected via the $\tau^+\to \pi^+\pi^0\bar \nu_\tau$ decay channel with the residual neutral showers and charged tracks.
The transition $\gamma(\pi^0)$ from the $D_s^{*+}$ and
the leptonic $D^+_s$ decay signals are distinguished from combinatorial backgrounds
by three kinematic variables
\begin{equation}
\Delta E \equiv \sqrt s-E_{\rm tag}-E_{\rm miss}-E_{\gamma(\pi^0)},\nonumber
\end{equation}
and

\begin{align}
\mathrm{MM}^{(*)2}&\equiv\left (\sqrt s-\Sigma_k E_k\right )^2
-|-\Sigma_k \vec{p}_k|^2.\nonumber
\end{align}
Here
$E_{\rm miss} \equiv \sqrt{|\vec{p}_{\rm miss}|^2+m_{D_s^+}^2}$ and
$\vec{p}_{\rm miss} \equiv -\vec{p}_{\rm tag}-\vec{p}_{\gamma(\pi^0)}$ are
the missing energy and momentum of the recoiling system of the transition $\gamma(\pi^0)$ and the single-tag $D_s^-$,
where $E_k$ and $\vec p_k$ are
the energy and momentum of the given particle $k$ $(\pi^+\pi^0$, transition $\gamma(\pi^0)$ or tag), respectively.
The MM$^{*2}$ and MM$^2$ are the missing masses squared of the signal $D^+_s$ and neutrinos, respectively. The index $k$ sums over the single-tag $D^-_s$ and the transition $\gamma(\pi^0)$ for MM$^{*2}$, while over the single-tag $D^-_s$, the transition $\gamma(\pi^0)$, and $\pi^+\pi^0$ for MM$^{2}$. Here, the MM$^{*2}$ is required to be within the interval (3.82,\,3.98)~GeV$^2/c^4$.
All remaining $\gamma$ and $\pi^0$ candidates
are looped over and the one giving the least $|\Delta E|$ is chosen as the transition $\gamma(\pi^0)$ candidate. The $\tau^+\to \pi^+\pi^0\bar{\nu}_\tau$ is actually dominated by $\tau^+\to \rho^+\bar \nu_\tau$. To form the $\rho^+$ candidate of the signal side, we use the same selection criteria as those of the tag side.
The charge of the pion candidate is required to be opposite to that of the single-tag $D^-_s$ meson.
To suppress the backgrounds with extra photon(s),
the sum of the energies deposited in the EMC of those unused showers in the double-tag event ($E_{\mathrm{extra}~\gamma}^{\rm sum}$) is required to be less than 0.1\,GeV based on an optimization
using the inclusive MC sample. Figure~\ref{fig:fig3}(a) shows the distribution of $E_{\rm extra~\gamma}^{\rm sum}$ of the double-tag candidates. The consistency between data and MC simulation around zero is not very good. The associated acceptance efficiency difference due to imperfect simulation will be corrected as discussed later.
Moreover, we require no extra good charged track in each event ($N_{\rm extra}^{\rm charge}=0$).

To check the quality of the reconstructed $\rho^+$, we examine the $M_{\pi^+\pi^0}$ spectrum
and the helicity angle of $\rho^+$ candidates ($\cos\theta_\rho$) of the selected double-tag candidates, as shown in
Figs.~\ref{fig:fig3}(b) and \ref{fig:fig3}(c).
The $\theta_\rho$ is calculated as an angle of the momentum of $\pi^+$ in the rest frame of $\rho^+$ with respect to the $\rho^+$ direction in the initial $e^+e^-$ beams, as the $\tau^+$ momentum is not available.
Figure~\ref{fig:mm2fit} shows the resulting ${\rm MM}^2$ distributions of the $D_s^+\to\tau^+\nu_\tau$ candidates selected from the data samples at various energy points.

\section{Branching fraction}
The efficiencies of reconstructing the double-tag  candidate events
are determined with exclusive signal MC samples of $e^+e^-\to D_s^+D^{*-}_s+c.c.$, where
the $D_s^-$ decays to each tag mode and the $D_s^+$ decays to $\tau^+\nu_\tau$ with $\tau^+\to\pi^+\pi^0\bar \nu_\tau$.
The double-tag efficiencies ($\epsilon^{i1}_{\rm DT}$) obtained at $\sqrt{s}=4.178$ GeV are summarized in the fourth column of Table~\ref{tab:bf}. The obtained $\epsilon^{j}_{\gamma(\pi^0)\tau^+\nu_\tau}$ at various energy points are summarized in the third column of Table~\ref{tab:eff2}.
These efficiencies have been corrected by a factor $f^{\rm cor}=1.058\times0.996\times0.991\times1.003$ to take into account the data-MC efficiency differences due to the requirements of
$E_{\mathrm{extra}~\gamma}^{\rm sum}$\&$N_{\rm extra}^{\rm charge}$, $\pi^+$ PID, $\rm MM^{*2}$, and the least $|\Delta E|$ as described in Sec.~\ref{sys}.

\begin{figure*}[htbp]
  \centering
  \includegraphics[width=1\textwidth]{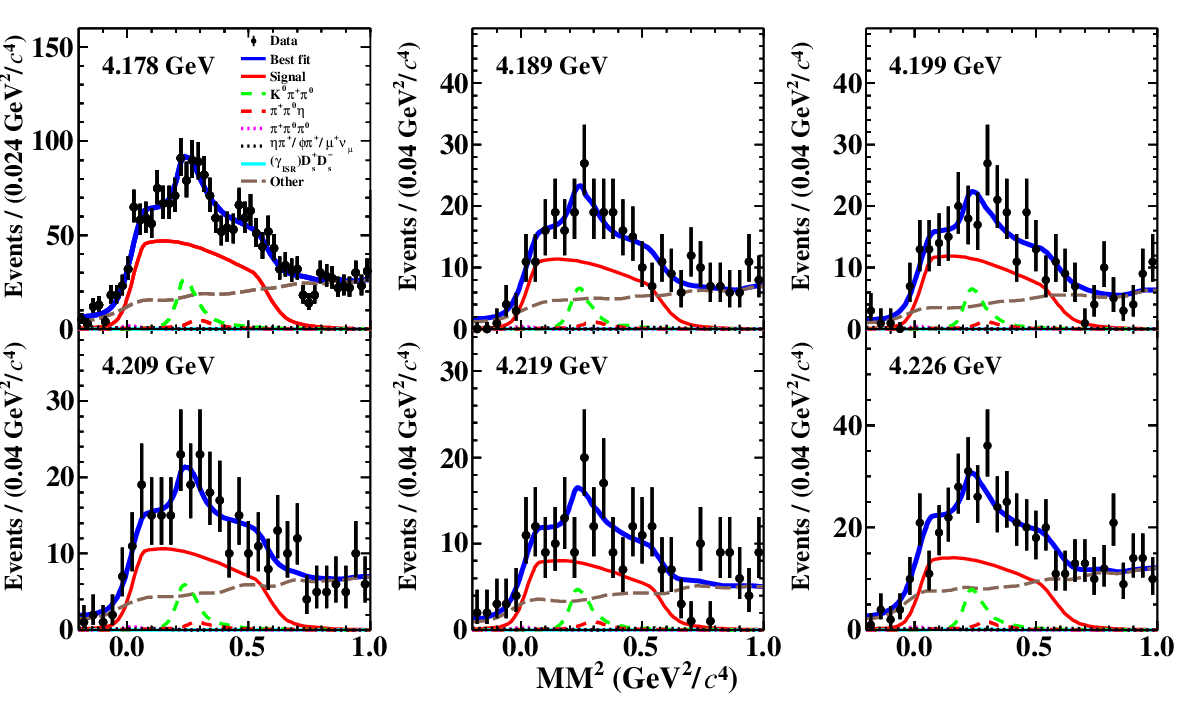}
  \caption{\footnotesize
Simultaneous fit to the ${\rm MM}^2$ distributions of the accepted $D_s^+\to\tau^+\nu_\tau$ candidates from the data samples at various energy points.
Points with error bars are data.
Solid blue curves are the fit results.
Red solid lines show the signals.
Green dashed, red dashed, pink dotted, black dotted, cyan solid, and brown dashed curves are the backgrounds from
$D_s^+\to K^0\pi^+\pi^0$,
$D_s^+\to\pi^+\pi^0\eta$,
$D_s^+\to\pi^+\pi^0\pi^0$,
$D^+_s\to (\eta\pi^+,\phi\pi^+,\mu^+\nu_\mu)$, $e^+e^-\to (\gamma_{\rm ISR})D_s^+D_s^-$,
and the other backgrounds after excluding the components aforementioned, respectively.
}
\label{fig:mm2fit}
\end{figure*}

To obtain the branching fraction for $D_s^+\to\tau^+\nu_\tau$, we perform a simultaneous fit to
the ${\rm MM^2}$ distributions, as shown in Fig.~\ref{fig:mm2fit}, where the six energy points are constrained to have a common leptonic decay branching fraction. For various energy points, the branching fractions are calculated by using Eq.~(\ref{eq1}) with $N^j_{\rm DT}$, $N^j_{\rm ST}$, and $\epsilon^j_{\gamma(\pi^0)\tau^+\nu_\tau}$. The shapes of the $D^+_s\to \tau^+\nu_\tau$ signals are described by a sum of two bifurcated-Gaussian functions, whose parameters are determined from the fits to the signal MC events and are fixed in the simultaneous fit. The peaking backgrounds of $D^+_s\to K^0\pi^+\pi^0$~\cite{arxiv210315098},
$D_s^+\to \pi^+\pi^0\pi^0$~\cite{PDG2020},
$D_s^+\to\pi^+\pi^0\eta$~\cite{bes3_etapipi0},
$D_s^+\to\eta \pi^+$~\cite{PDG2020}, $D_s^+\to\phi\pi^+$~\cite{PDG2020}, and $D_s^+\to\mu^+\nu_\mu$~\cite{bes2019} are modeled by the corresponding simulated shapes.
The $D_s^+\to\pi^+\pi^0\eta$ decays are generated using the amplitude-analysis results in Ref.~\cite{bes3_etapipi0}.
The $D_s^+\to\eta\pi^+$, $D_s^+\to\phi\pi^+$, and $D_s^+\to\mu^+\nu_\mu$ decays are uniformly generated across the event phase space.
To model the resonant contributions in the $D^+_s\to K^0\pi^+\pi^0$ and $D_s^+\to \pi^+\pi^0\pi^0$ decays,
these two decays are generated with a modified data-driven generator BODY3~\cite{ref:evtgen,etaX}, which was developed to simulate different intermediate states in data for a given three-body final state. The two-dimensional distributions of $M_{K^0\pi^+}^2$ versus $M_{\pi^+\pi^0}^2$ and $M_{\pi^+\pi^0}^2$ versus $M^2_{\pi^0\pi^0}$ found in data, corrected for backgrounds and efficiencies, are taken as the input for the BODY3 generator. The efficiencies across the kinematic space are obtained with the MC samples generated with the modified phase-space generator.
For $D^+_s\to K^0\pi^+\pi^0$,
the interaction between the $K_L^0$ particle and the EMC materials may not be well simulated, thus causing
large difference between the acceptance efficiency of data and that of simulation due to the requirement of $E_{\rm extra \gamma}^{\rm sum}<0.1\,\rm GeV$.
Therefore, the sizes of the $D_s^+\to K^0\pi^+\pi^0$ background are float, but their rates over the simulated ones at the six energy points are constrained to be the same. The yields of the peaking backgrounds of $D_s^+\to \pi^+\pi^0\pi^0$,
$D_s^+\to\pi^+\pi^0\eta$,
$D_s^+\to\eta\pi^+$, $D_s^+\to\phi\pi^+$, and $D_s^+\to\mu^+\nu_\mu$ are estimated based on the MC simulated misidentification efficiencies and the world average branching fractions, and their sizes are fixed in the fit.
The simulated shapes of these peaking backgrounds have been smeared with a Gaussian function, with parameters obtained from the control sample of $D^+_s\to \eta\rho^+$.
The background of $D_s^-\to$ tags versus $D_s^+\to$ signals from $e^+e^-\to (\gamma_{\rm ISR})D_s^+D_s^-$ contributes about 0.3\% of the observed signal yield and its relative ratio is also fixed in the fit.
The other combinatorial backgrounds
are modeled by the shapes from the inclusive MC sample after excluding the components aforementioned.

The simultaneous fit results are also shown in Fig.~\ref{fig:mm2fit}.
From this fit, the branching fraction for $D^+_s\to \tau^+\nu_\tau$ is obtained to be $(5.29\pm0.25)\%$. This corresponds to the signal yield of $D^+_s\to \tau^+\nu_\tau$ to be $1745\pm 84$, where the uncertainty is statistical only.

\section{Systematic uncertainties}
\label{sys}

With the DT method, most of uncertainties related to the single-tag selection are canceled.
Sources of the systematic uncertainties in the branching fraction measurement are summarized in Table~\ref{tab:sys_tot}.
Each of them, which is estimated relative to the measured branching fraction, is described below.

\begin{table}[htbp]\centering
\caption{\footnotesize Systematic uncertainties in the branching fraction measurement.}
\small
\begin{tabular}{cc}
\hline
Source & Uncertainty\,(\%)\\ \hline
Single-tag yield                                                                                          &0.6\\
$\pi^+$ tracking                                                                                                      &0.2\\
$\pi^+$ PID                                                                                                             &0.2\\
$\gamma\,(\pi^0)$ reconstruction                                                                              &2.1\\
$E_{\mathrm{extra}~\gamma}^{\rm sum}$ and $N_{\rm extra}^{\rm charge}$ requirements   &2.2\\
MM$^{*2}$ requirement                                                                                             &0.8\\
$\tau^+$ decay & 1.2\\
$\rm MM^2$ fit                                                                                                        &1.3\\
Least $|\Delta E|$                                                                                                     &0.4\\
Tag bias                                                                                                                  &0.5\\
MC statistics                                                                                                           &0.3\\
Quoted branching fractions                                                                                       &0.5\\
\hline
Total                                                                                                                       &3.8\\
\hline
\end{tabular}
\label{tab:sys_tot}
\end{table}

\subsection{Determination of single-tag yield}
The uncertainty in the total number of the single-tag $D_s^-$ mesons is assigned to be 0.6\% by taking into account the background fluctuation in the fit, and examining the changes
of the fit yields when varying the signal shape, background shape.

\subsection{$\pi^+$ tracking and PID}
The $\pi^+$ tracking and PID efficiencies are studied with the $e^+e^-\to K^+K^-\pi^+\pi^-$ events.
The data-MC efficiency ratios of the $\pi^+$ tracking and PID efficiencies are $1.000\pm0.002$ and $0.996 \pm 0.002$, respectively. After multiplying the signal efficiencies by the latter factor, we assign 0.2\% and 0.2\%
as the systematic uncertainties arising from the $\pi^+$ tracking and PID efficiencies, respectively.

\subsection{$\gamma(\pi^0)$ reconstruction}
The photon selection efficiency was previously studied with the $J/\psi\to\pi^+\pi^-\pi^0$ decays~\cite{geff}.
The $\pi^0$ reconstruction efficiency was previously studied with the $e^+e^-\to K^+K^-\pi^+\pi^-\pi^0$ events.
The systematic uncertainty of finding the transition $\gamma(\pi^0)$, which is weighted according to the branching fractions for $D_s^{*+}\to\gamma D_s^+$ and $D_s^{*+}\to\pi^0D_s^+$~\cite{PDG2020}, is obtained to be 1.0\%. For the $\pi^0$ in the leptonic decay, the relevant systematic uncertainty is assigned to be 1.1\%. The total systematic uncertainty related to the photon and $\pi^0$ reconstruction is obtained to be 2.1\% by adding these two uncertainties linearly.

\subsection{$E_{\mathrm{extra}~\gamma}^{\rm sum}$ and $N_{\rm extra}^{\rm charge}$ requirements}
The efficiencies for the combined requirements of $E_{\mathrm{extra}~\gamma}^{\rm sum}$ and
 $N_{\rm extra}^{\rm charge}$
are investigated with the double-tag sample of $D^+_s\to \eta \pi^+$,
which has similar acceptance efficiencies to our signals.
The ratio of the averaged efficiency of data to that of simulation is $1.058\pm0.022$. After multiplying the signal efficiency by this factor, we assign 2.2\% as the relevant systematic uncertainty.

\subsection{${\rm MM}^{*2}$ requirement}
\label{secE}
To assign the systematic uncertainty originating from the ${\rm MM}^{*2}$ requirement, we fit to the ${\rm MM}^{*2}$ distribution of the accepted $D^+\to \tau^+\nu_\tau$ candidates in data after excluding this requirement.
In the fit, the background shape is derived from the inclusive MC sample and the signal shape is described by the shape from the
signal MC events convolved with a Gaussian function to take into account the difference between data and simulation.
The parameters of the Gaussian function are floated.
From the fit, the mean and resolution of the Gaussian function are obtained to be 0.008~GeV$^2$/$c^4$ and 0.012~GeV$^2$/$c^4$ respectively. Then
we examine the signal efficiency after smearing the corresponding Gaussian function to the ${\rm MM}^{*2}$ variable.
The ratio of the acceptance efficiencies with and without the smearing is $0.991\pm0.008$.
After multiplying the signal efficiency by the factor, we assign 0.8\% as the systematic uncertainty of the ${\rm MM}^{*2}$ requirement.

\subsection{$\tau^+$ decay}
 The difference of the measured branching fractions with and without taking into account  $\tau^+\to (\pi^+\pi^0)_{\rm non{\text -}\rho}\bar \nu_\tau$~\cite{PDG2020}, 1.2\%, is considered as a systematic uncertainty. The uncertainty due to imperfect simulation of the $M_{\pi^+\pi^0}$ lineshape is assigned with the same method described in Sec.~\ref{secE}. From the fit to the $M_{\pi^+\pi^0}$ distribution of data,
the mean and resolution of the Gaussian function used to smear the $M_{\pi^+\pi^0}$ distribution are obtained to be (0.010, 0.008)~GeV/$c^2$. The difference of the signal efficiencies with and without smearing is negligible.

\subsection{${\rm MM}^2$ fit}
The systematic uncertainty in the ${\rm MM}^2$ fit is considered in three aspects. At first, we vary the estimated yields of peaking backgrounds from $D^+_s\to K^0\pi^+\pi^0$~\cite{arxiv210315098},
$D_s^+\to \pi^+\pi^0\pi^0$~\cite{PDG2020},
$D_s^+\to\pi^+\pi^0\eta$~\cite{bes3_etapipi0},
$D_s^+\to\eta \pi^+$~\cite{PDG2020}, $D_s^+\to\phi\pi^+$~\cite{PDG2020}, and $D_s^+\to\mu^+\nu_\mu$~\cite{bes2019} by $\pm1\sigma$ of the quoted branching fractions and the input cross section~\cite{crosssection}. Then, we vary the peaking background yields of $D^+_s\to \pi^+\pi^0\eta$ and $D^+_s\to \pi^+\pi^0\pi^0$ by $-20\%$, based on the data-MC difference of the in-efficiency of photon(s). Finally, we float the parameters of two bifurcated-Gaussian functions and the convoluted Gaussian functions by $\pm1\sigma$. The quadratic sum of the relative changes of the re-measured branching fractions, 1.3\%, is assigned as the corresponding systematic uncertainty.

\subsection{Selection of the transition $\gamma\,(\pi^0)$ with the least $|\Delta E|$}
The systematic uncertainty from the selection of the transition $\gamma\,(\pi^0)$ from $D_s^{*+}$ with the least $|\Delta E|$ method is estimated by using the control samples of $D^+_s\to K^+K^-\pi^+$ and $D^+_s\to \eta\pi^0\pi^+$. The ratio of the efficiency of selecting the transition $\gamma\,(\pi^0)$ candidates of data to that in simulation is $1.003\pm0.004$. After multiplying the signal efficiency by this factor, we take 0.4\% as the corresponding systematic uncertainty.

\subsection{Tag bias}
The single-tag efficiencies in the inclusive and signal MC samples may be slightly different from each other due to different track multiplicities in these two environments. This may cause incomplete cancelation of the uncertainties of the single-tag selection efficiencies. The associated uncertainty is assigned as 0.5\%, by taking into account the differences of the tracking and PID efficiencies of $K^\pm$ and $\pi^\pm$ as well as the selections of neutral particles between data and simulation in different environments.

\subsection{MC statistics}
The uncertainty due to the finite MC statistics 0.3\%, which is dominated by that of the double-tag efficiency, is considered as a source of systematic uncertainty.

\subsection{Quoted branching fractions}
The uncertainties of the quoted branching fractions for $\pi^0\to\gamma\gamma$ and $\tau^+\to\pi^+\pi^0\bar \nu_\tau$ are 0.03\% and 0.4\%, respectively.
The world average branching fractions for $D_s^{*-}\to\gamma D_s^-$ and $D_s^{*-}\to\pi^0 D_s^-$ are $(93.5\pm0.7)\%$ and $(5.8\pm0.7)\%$, respectively, which are fully correlated with each other. An associated uncertainty is assigned by re-weighting $\varepsilon_{\gamma\tau^+\nu_\tau}$ and
$\varepsilon_{\pi^0\tau^+\nu_\tau}$ via varying these two branching fractions by $\pm 1\sigma$.
The change of the re-weighted signal efficiency is 0.2\%. The uncertainty of the branching fraction for $D^{*-}\to e^+e^-D^-_s$, 0.2\%, is considered as an additional uncertainty.
The total systematic uncertainty associated with the above branching fractions is obtained to be 0.5\%,
by adding these four uncertainties in quadrature.

\subsection{Total systematic uncertainty}
The total systematic uncertainty in the measurement of the branching fraction for $D^+_s\to \tau^+\nu_\tau$ is determined to be 3.8\% by adding all above uncertainties in quadrature.

\section{Results}

Combining our branching fraction

\begin{equation*}
{\mathcal B}_{D_s^+\to\tau^+\nu_\tau}=(5.29\pm0.25_{\rm stat}\pm0.20_{\rm syst})\%
\end{equation*}
and the world average values of $G_F$, $m_\mu$, $m_{D^+_s}$, and $\tau_{D^+_s}$~\cite{PDG2020} in Eq.~(1) with $\Gamma_{D_s^+\to\tau^+\nu_\tau}=\mathcal B_{D_s^+\to\tau^+\nu_\tau}/\tau_{D_s^+}$ yields

\begin{equation*}
f_{D_s^+}|V_{cs}|=(244.8\pm5.8_{\rm stat}\pm4.8_{\rm syst})~\mathrm{MeV}.
\end{equation*}
Here the systematic uncertainties arise mainly from the uncertainties in the measured
branching fraction (3.8\%) and the $D^+_s$ lifetime (0.8\%).
Taking $|V_{cs}|=0.97320\pm0.00011$ from the global fit
in the standard model~\cite{PDG2020, ckmfitter}, we obtain $f_{D_s^+}=(251.6\pm5.9_{\rm stat}\pm4.9_{\rm syst})~\mathrm{MeV}$. Alternatively, taking
$f_{D_s^+}=(249.9\pm0.5)~\mathrm{MeV}$ of the recent LQCD calculations~\cite{FLab2018,LQCD,etm2015,FLAG2019}
as input, we determine $|V_{cs}|=0.980\pm0.023_{\rm stat}\pm0.019_{\rm syst}$.
One additional systematic uncertainty of the input $f_{D_s^+}$
is 0.2\%, while that of $|V_{cs}|$ is negligible.
The $|V_{cs}|$ measured in this work is in agreement with our measurements via the
$D\to\bar K\ell^+\nu_\ell$ decays~\cite{bes3_kev,bes3_ksev,bes3_klev,bes3_kmuv}, the $D_s^+\to\mu^+\nu_\mu$ decay~\cite{bes2019}, and the $D_s^+\to\eta^{(\prime)}e^+\nu_e$ decays~\cite{bes3_etaev}.

Using the branching fraction of $\mathcal{B}_{D_s^+\to\mu^+\nu_\mu}=(5.35\pm0.21)\times10^{-3}$~\cite{hajime2021},
$\mathcal R_{\tau/\mu}$ is determined to be
$9.89\pm0.71,$
which agrees with the standard model predicted value of $9.75\pm0.01$ within 1$\sigma$.

\section{Summary}
By analyzing $6.32~\mathrm{fb}^{-1}$ of $e^+e^-$ collision data collected between
4.178 and 4.226 GeV with the BESIII detector, we present a measurement of $D_s^+\to\tau^+\nu_\tau$
using the $\tau^+\to\pi^+\pi^0\bar \nu_\tau$ decay channel.
The branching fraction for $D_s^+\to\tau^+\nu_\tau$ is determined to be $(5.29\pm0.25\pm0.20)\%$,
which is well consistent with previous measurements~\cite{PDG2020}.
Combining this branching fraction with the $|V_{cs}|$ given by CKMfitter~\cite{PDG2020, ckmfitter},
we obtain $f_{D^+_s}=(251.6\pm5.9\pm4.9)$~MeV.
Conversely, combining this branching fraction with the $f_{D^+_s}$ calculated by the latest LQCD~\cite{FLab2018,LQCD,etm2015,FLAG2019},
we determine $|V_{cs}|=0.980\pm0.023\pm0.019$.
Combining our branching fraction with $\mathcal B(D^+_s \to \mu^+\nu_\mu)=(5.35\pm0.21)\times10^{-3}$~\cite{hajime2021}, we determine $\mathcal R_{\tau/\mu}=9.89\pm0.71$, which is consistent with the expectation based on lepton flavor universality. This ratio implies that no lepton flavor universality violation is found between the $D_s^+ \to \tau^+ \nu_\tau$ and  $D_s^+ \to \mu^+ \nu_\mu$ decays under the current precision.
Combining our branching fraction with the one measured via $\tau^+\to\pi^+\bar\nu_\tau$~\cite{hajime2021}, we obtain $\mathcal B(D_s^+\to\tau^+\nu_\tau)=(5.24\pm0.18\pm0.14)\%$, $f_{D^+_s}=(250.4\pm4.3\pm3.4)$~MeV, $|V_{cs}|=0.975\pm0.017\pm0.013$, and $\mathcal R_{\tau/\mu}=9.79\pm0.57$, where the uncertainties from the single-tag yield, the $\pi^\pm$ tracking efficiency, the soft $\gamma$ reconstruction, the best transition photon selection, and the tag bias are treated to be fully correlated for $\mathcal B(D_s^+\to\tau^+\nu_\tau)$, additional common uncertainties come from $\tau_{D_s^+}$, $m_{D^+_s}$, and $m_\tau$ for $f_{D^+_s}$ and $|V_{cs}|$,
and all the other uncertainties are independent.

\section{Acknowledgement}

The BESIII collaboration thanks the staff of BEPCII and the IHEP computing center for their strong support. This work is supported in part by National Key R\&D Program of China under Contracts Nos. 2020YFA0406400, 2020YFA0406300; National Natural Science Foundation of China (NSFC) under Contracts Nos. 11775230, 11805037, 11625523, 11635010, 11735014, 11822506, 11835012, 11935015, 11935016, 11935018, 11961141012, 12022510, 12025502, 12035009, 12035013, 12061131003; the Chinese Academy of Sciences (CAS) Large-Scale Scientific Facility Program; Joint Large-Scale Scientific Facility Funds of the NSFC and CAS under Contracts Nos. U1832121, U1732263, U1832207; CAS Key Research Program of Frontier Sciences under Contract No. QYZDJ-SSW-SLH040; 100 Talents Program of CAS; INPAC and Shanghai Key Laboratory for Particle Physics and Cosmology; ERC under Contract No. 758462; European Union Horizon 2020 research and innovation programme under Contract No. Marie Sklodowska-Curie grant agreement No 894790; German Research Foundation DFG under Contracts Nos. 443159800, Collaborative Research Center CRC 1044, FOR 2359, FOR 2359, GRK 214; Istituto Nazionale di Fisica Nucleare, Italy; Ministry of Development of Turkey under Contract No. DPT2006K-120470; National Science and Technology fund; Olle Engkvist Foundation under Contract No. 200-0605; STFC (United Kingdom); The Knut and Alice Wallenberg Foundation (Sweden) under Contract No. 2016.0157; The Royal Society, UK under Contracts Nos. DH140054, DH160214; The Swedish Research Council; U. S. Department of Energy under Contracts Nos. DE-FG02-05ER41374, DE-SC-0012069.

\end{document}

%% file: BESIII_author.tex
\author{M.~Ablikim$^{1}$, M.~N.~Achasov$^{10,b}$, P.~Adlarson$^{67}$, S. ~Ahmed$^{15}$, M.~Albrecht$^{4}$, R.~Aliberti$^{28}$, A.~Amoroso$^{66A,66C}$, M.~R.~An$^{32}$, Q.~An$^{63,49}$, X.~H.~Bai$^{57}$
% ...
\\
\vspace{0.2cm}
(BESIII Collaboration)\\
\vspace{0.2cm} {\it
$^{1}$ Institute of High Energy Physics, Beijing 100049, People's Republic of China\\
% ...
$^{4}$ Bochum Ruhr-University, D-44780 Bochum, Germany\\
% ...
$^{10}$ G.I. Budker Institute of Nuclear Physics SB RAS (BINP), Novosibirsk 630090, Russia\\
% ...
$^{15}$ Helmholtz Institute Mainz, Staudinger Weg 18, D-55099 Mainz, Germany\\
% ...
$^{28}$ Johannes Gutenberg University of Mainz, Johann-Joachim-Becher-Weg 45, D-55099 Mainz, Germany\\
% ...
$^{32}$ Liaoning Normal University, Dalian 116029, People's Republic of China\\
% ...
$^{49}$ State Key Laboratory of Particle Detection and Electronics, Beijing 100049, Hefei 230026, People's Republic of China\\
% ...
$^{57}$ University of Jinan, Jinan 250022, People's Republic of China\\
% ...
$^{63}$ University of Science and Technology of China, Hefei 230026, People's Republic of China\\
% ...
$^{66}$ University of Turin and INFN, (A)University of Turin, I-10125, Turin, Italy; (B)University of Eastern Piedmont, I-15121, Alessandria, Italy; (C)INFN, I-10125, Turin, Italy\\
$^{67}$ Uppsala University, Box 516, SE-75120 Uppsala, Sweden\\
% ...
$^{b}$ Also at the Novosibirsk State University, Novosibirsk, 630090, Russia\\
% ...
}
}